\documentclass[aps,pre,twocolumn,superscriptaddress]{revtex4-2}
\usepackage{nameref}
\usepackage{amssymb}
\usepackage{amsthm}
\usepackage{amsmath}
\usepackage{mathrsfs}
\usepackage{graphicx}
\usepackage{color}
\usepackage{physics}
\usepackage{bm}
\usepackage{booktabs}
\usepackage{subfigure}
\usepackage{float}
\usepackage[colorlinks,urlcolor=blue,linkcolor=blue,anchorcolor=blue,citecolor=blue]{hyperref}
\usepackage{lineno}

\begin{document}
\title{Impact of on-site potentials on $q$-breathers in nonlinear chains}

\author{Lin Deng}
\affiliation{Lanzhou Center for Theoretical Physics, Key Laboratory of Theoretical Physics of Gansu Province, Key Laboratory of Quantum Theory and Applications of MoE, Gansu Provincial Research Center for Basic Disciplines of Quantum Physics, Lanzhou University, Lanzhou, Gansu 730000, China}

\author{Chengguan Fang}
\affiliation{Lanzhou Center for Theoretical Physics, Key Laboratory of Theoretical Physics of Gansu Province, Key Laboratory of Quantum Theory and Applications of MoE, Gansu Provincial Research Center for Basic Disciplines of Quantum Physics, Lanzhou University, Lanzhou, Gansu 730000, China}

\author{Hang Yu}
\affiliation{Lanzhou Center for Theoretical Physics, Key Laboratory of Theoretical Physics of Gansu Province, Key Laboratory of Quantum Theory and Applications of MoE, Gansu Provincial Research Center for Basic Disciplines of Quantum Physics, Lanzhou University, Lanzhou, Gansu 730000, China}

\author{Yisen Wang}\email{wys@lzu.edu.cn}
\affiliation{Lanzhou Center for Theoretical Physics, Key Laboratory of Theoretical Physics of Gansu Province, Key Laboratory of Quantum Theory and Applications of MoE, Gansu Provincial Research Center for Basic Disciplines of Quantum Physics, Lanzhou University, Lanzhou, Gansu 730000, China}

\author{Zhigang Zhu}
\affiliation{Department of Physics, Lanzhou University of Technology, Lanzhou, Gansu 730000, China}
\affiliation{Lanzhou Center for Theoretical Physics, Key Laboratory of Theoretical Physics of Gansu Province, Key Laboratory of Quantum Theory and Applications of MoE, Gansu Provincial Research Center for Basic Disciplines of Quantum Physics, Lanzhou University, Lanzhou, Gansu 730000, China}

\author{Weicheng Fu}
\affiliation{Department of Physics, Tianshui Normal University, Tianshui 741001, Gansu, China}
\affiliation{Lanzhou Center for Theoretical Physics, Key Laboratory of Theoretical Physics of Gansu Province, Key Laboratory of Quantum Theory and Applications of MoE, Gansu Provincial Research Center for Basic Disciplines of Quantum Physics, Lanzhou University, Lanzhou, Gansu 730000, China}

\author{Sergej Flach}\email{sflach@ibs.re.kr}
\affiliation{Center for Theoretical Physics of Complex Systems, Institute for Basic Science, Daejeon 34126, Republic of Korea}
\affiliation{Basic Science Program, Korea University of Science and Technology, Daejeon 34113, Republic of Korea}

\author{Liang Huang}\email{huangl@lzu.edu.cn}
\affiliation{Lanzhou Center for Theoretical Physics, Key Laboratory of Theoretical Physics of Gansu Province, Key Laboratory of Quantum Theory and Applications of MoE, Gansu Provincial Research Center for Basic Disciplines of Quantum Physics, Lanzhou University, Lanzhou, Gansu 730000, China}

\begin{abstract}
On-site potentials are ubiquitous in physical systems and strongly influence their heat transport and energy localization. These potentials will inevitably affect the dynamical properties of $q$-breathers (QBs), defined as periodic orbits exponentially localized in normal mode space.
By integrating on-site terms into the Fermi-Pasta-Ulam-Tsingou-$\beta$ system, this work utilizes numerical simulations and Floquet analysis to systematically explore the influence of on-site potentials on QB stability.
For most QBs, except those at the phonon band edges, the instability is primarily governed by parametric resonance, and effectively described by coupled Mathieu equations. This approach provides a theoretical expression for the instability thresholds, which aligns well with numerical results. We demonstrate that the instability thresholds can be controlled through the strength of on-site potentials, and for a strong enough quadratic on-site potential, the QBs are always stable.  Furthermore, the instability threshold is highly sensitive to the seed mode, in stark contrast to systems without on-site potentials. In addition, the instability phase diagrams exhibit joint interplay between different terms in the Hamiltonian, such as the quadratic on-site and quartic inter-site interaction terms, in regulating the QB dynamics. These findings offer valuable insights into QB stability and the manipulation of localized excitations in diverse physical systems with on-site potentials.
\end{abstract}

\maketitle

\section{Introduction}
On-site potentials, or substrate potentials, are a key concept in condensed matter physics and nonlinear dynamics, playing a central role in the behavior of discrete lattices and extended systems, such as ultra-cold atomic gases loaded into optical potentials \cite{Singh2012PhysRevA, He2018PhysRevA}, light propagation through photonic crystals \cite{joannopoulos_photonic_2008}, and respective theoretical modeling of networks of nonlinear oscillators \cite{Shiroky2018PhysRevE, Gadasi2022PhysRevLett, Gershenzon2023PhysRevLett}, and the Bose-Hubbard model \cite{Jaksch1998PhysRevLett}, among others. Mathematically, it is typically incorporated as an additive term in the system's Hamiltonian, modifying the energy landscape at specific sites \cite{Remoissenet1984PhysRevB,Yem2002PhysRevE,Yem2005PhysRevB,Woulach2013PhysRevE,BUCHHEIT2020PhysicaD}. These potentials serve as useful models for a range of realistic scenarios, such as impurities in crystalline solids \cite{Gehrmann2019NC}, externally applied fields \cite{Chinellato2024PRB}, and local trapping potentials in cold atom systems \cite{Potirniche2017PhysRevLett,Tomza2019RevModPhys}, which result in diverse phenomena \cite{Aoki2016PhysRevE,KIVSHAR1993PRA,Pettini1991PhysRevA,Pistone2018EL,Danieli2019PhysRevE,YAO2021AOP}.
For instance, in the Bose-Hubbard model, an increase in the ratio of on-site potential to tunneling strength leads to a superfluid-to-Mott-insulator transition \cite{Jaksch1998PhysRevLett, Fisher1989PhysRevB}, and stronger three-body on-site interactions expand the Mott-insulating region \cite{Chen2008PhysRevA,Gimperlein2005PhysRevLett}.
In the realm of non-equilibrium statistical physics, on-site potentials significantly influence the heat transport behaviors of nonlinear systems \cite{ZhaoHong2000PhysRevE,Savin2003PhysRevE, Savin2004PhysRevLett}, and diffusive energy transport adhering to Fourier’s law was observed in models like ``ding-a-ling'' \cite{Casati1984PhysRevLett,ZhaoHong1998PhysRevE}. Nonlinear on-site potentials also lead to phenomena like negative differential resistance, anomalous diffusion, and rectification \cite{HuBambi2011PhysRevE, Luo2020PhysRevE}. Additionally, soliton dynamics and depinning forces are strongly influenced by substrate potential shapes \cite{Peyrard1982PhysRevB, HuBambi2005PhysRevE}.

Energy localization in systems with on-site potentials presents intriguing dynamics and has drawn significant interest. Defined as an external potential acting locally on individual sites, the on-site potential introduces spatial inhomogeneity that alters the dynamics and stability of localized modes like discrete breathers (DBs) \cite{Flach:1998,Campbell2004PhysicsToday,Flach:2008}. DBs, also referred to as intrinsic localized modes, are time-periodic oscillations characterized by spatially exponential localization of the energy density distribution\cite{ma94n,f94pre,f95preII}. DB solutions, commonly surviving in a strong nonlinear regime, have been studied in diverse contexts, including discrete nonlinear Schr\"odinger equations with alternating on-site potentials \cite{Johansson2004PhysRevE} and Klein-Gordon chains \cite{Gorbach2003PhysRevE}. On-site potentials underpin various energy localization phenomena \cite{PEYRARD1998PhysicaD}, including gap DBs \cite{Gorbach2003PhysRevE} and moving breather collisions \cite{Alvarez2008PLA}. In the nonlinear chain with sixth-order polynomial on-site potentials, DBs can affect the specific heat of the system \cite{SingXiong2021}.

To characterize energy localization in normal mode space, $q$-breathers (QBs) are proposed as an analogy to DBs \cite{Flach2005PhysRevLett}, which are periodic orbits composed of a small fraction of normal modes, with energy exponentially localized around these modes \cite{Flach2006PhysRevE}. Slight perturbations of QBs thereby approximately form low-dimensional tori within the phase space.
This concept provides crucial insights into the Fermi-Pasta-Ulam-Tsingou (FPUT) recurrence, which declares that the state in phase space spanned by normal coordinates returns close to its initial configuration in a quasi-periodic manner, rather than equilibrating as initially expected \cite{Flach2006PhysRevE, FLACH2008PhysicaD}.
QBs have been validated in various systems, including FPUT-$\beta$ lattices \cite{Ivanchenko2006PhysRevLett}, Bose-Hubbard chains \cite{Nguenang2007PhysRevB, Pinto2009PhysicaD}. QBs also survive in disordered systems, with their stability depending on the details of disorder \cite{Ivanchenko2009PhysRevLett}.
The existence of QBs in systems with on-site potential has been analyzed within the framework of the discrete nonlinear Schr\"odinger model \cite{Mishagin2008NJP}.

The stability of QBs is a central issue in understanding their role in the dynamics and thermalization of nonlinear systems \cite{Flach2006PhysRevE}.
For weak nonlinearity, QBs are stable and isolated from energy exchange with other modes. However, as the nonlinearity strength increases, instabilities of QBs emerge due to mechanisms such as parametric resonance and Chirikov resonances \cite{Yoshimura2000PhysRevE, CHIRIKOV1979PhysRep}.
The instability thresholds can be obtained from bifurcation points in the multipliers spectrum of a Floquet analysis, which aligns with the onset of weak chaos \cite{Flach2005PhysRevLett,DeLuca1995Chaos}. In the FPUT system, a notable feature is that the instability threshold is independent of the choice of the seed mode used to generate the QBs \cite{Flach2005PhysRevLett,Flach2006PhysRevE}.
As QBs become unstable, the energy initially localized in the QBs gradually spreads to other modes,
facilitating energy redistribution and ultimately leading to thermalization \cite{wang2018pre}.
Despite significant advancements, a more systematic understanding of the effects of on-site potentials on QB stability is still lacking. Moreover, the combined influence of on-site potentials and nonlinear inter-site interactions on the QB instability dynamics remains unexplored.

In this paper, we systematically investigate the impact of on-site potentials on the QBs dynamics in nonlinear chains.
Section \ref{sec2} introduces the model by incorporating on-site potentials into the FPUT-$\beta$ Hamiltonian, allowing it to better represent realistic systems, such as the low-dimensional crystal subject to external fields.
Section \ref{sec3} presents results for the case where the nonlinear inter-site potential is absent. Specifically, QB solutions are obtained by solving for the roots of the Poincar{\'e} map with a generalized Newton's method. A detailed stability analysis based on Floquet theory reveals the combined effects of the quadratic and quartic on-site potential coefficients on QB stability, along with an examination of the associated modal dynamics.
In Sec. \ref{sec:betaOnsite}, we extend the analysis to systems with nonlinear inter-site potentials, offering deeper insights into how on-site and inter-site nonlinearities jointly regulate the instability thresholds.
Finally, a concise summary and discussion are provided in Sec. \ref{sec:Conclusion}.
Our results highlight the crucial role of on-site potentials in engineering localized energy states, providing valuable theoretical insights for the design of robust systems with well-defined localized modes.

\section{Model and Method}\label{sec2}

\subsection{Model}
\begin{figure}
    \centering
    \includegraphics[width=1\linewidth]{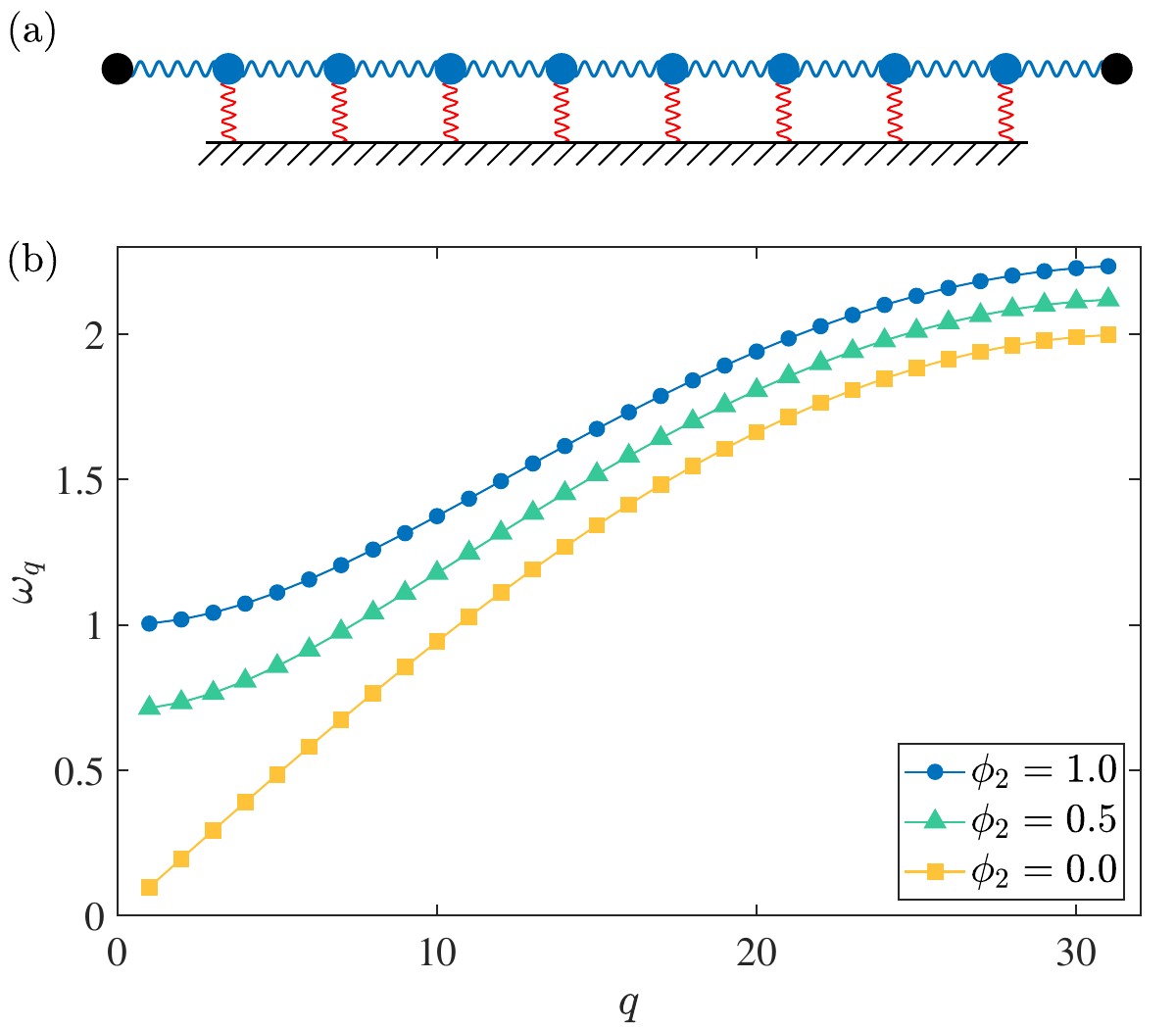}
    \caption{(a) Schematic diagram of the anharmonic chain with on-site potential. The red springs depict the on-site potential contributions, while the blue springs represent inter-site interactions. (b) Dispersion relations for the system for different values of the on-site potential strength $\phi_2$.}
    \label{Fig1_dispersion}
\end{figure}

We consider a nonlinear lattice consisting of $N$ identical particles, as schematically shown in Fig. \ref{Fig1_dispersion} (a). The Hamiltonian
\begin{equation}
\begin{split}
  H & =\sum_{i=1}^{N} \frac{p_i^2}{2} + \sum_{i=0}^{N} \left[\frac 12 \left(x_{i+1}-x_i\right)^2+\frac {\beta}4\left(x_{i+1}-x_i\right)^4 \right]\\
     & + \sum_{i=1}^{N} \left(\frac{\phi_2}{2} x_i^2 + \frac{\phi_4}{4} x_i^4\right)
\end{split}
\end{equation}
where $x_i$ represents the displacement of the $i$th particle from its equilibrium position, and $p_i$ is the conjugate momentum. The coefficient $\beta$ governs the strength of the quartic inter-site potential, while $\phi_2$ and $\phi_4$ control the quadratic and quartic on-site potentials, respectively.
This model extends the classical FPUT-$\beta$ chain by adding on-site potentials, making it applicable to more realistic scenarios where particles experience external forces, thus introducing additional complexity to the system's dynamics.

In this model, multiple potential terms can affect the stability of the QBs. To systematically investigate the influence of the parameters $\beta$, $\phi_2$, and $\phi_4$ on QB stability, we sequentially set each parameter to zero and analyze the effects of the remaining two parameters. We specifically consider three cases:
\begin{itemize}
\item Case I: $\beta = 0$, with $\phi_2$ and $\phi_4$ nonzero, where the harmonic chain has a gapped (away from zero) optical frequency spectrum, and nonlinearity arises solely from the on-site potential. Then the system is effectively a discrete Klein-Gordon lattice.
\item Case II: $\phi_4 = 0$, with $\beta$ and $\phi_2$ nonzero, which involves both quartic inter-site and quadratic on-site potentials. The harmonic chain is identical to Case I.
\item Case III: $\phi_2 = 0$, with $\beta$ and $\phi_4$ nonzero, where the system is influenced by both quartic inter-site and on-site potentials. The harmonic chain has a gapless acoustic frequency spectrum.
\end{itemize}
These cases allow us to systematically study how each potential term and their interplay affect QB stability.

The quadratic on-site potential modifies the dispersion relation of the system. By imposing fixed boundary conditions, $x_0 = x_{N+1} =p_0=p_{N+1}= 0$, the dispersion relation for the linearized system is derived as
\begin{equation}
\omega_q = \sqrt{\phi_2 + \Omega_q^2}, \quad \Omega_q = 2 \sin\left(\frac{q\pi}{2(N+1)}\right),
\end{equation}
where $q = 1, 2, \dots, N$ denotes the mode index, and $\Omega_q$ corresponds to the dispersion relation for a harmonic chain without on-site potential.
Figure $\ref{Fig1_dispersion}$(b) shows the dispersion relation of the systems with varying quadratic on-site parameters $\phi_2$. The on-site potential introduces a upward shift in the frequency spectrum, raising the lower bound of the phonon band to a size-independent value $\sqrt{\phi_2}$. Consequently, the frequencies of all normal modes increase and are gapped away from zero, resulting in stiffer oscillations.
This effect has profound implications for the system's dynamics, which, for instance, facilitates the existence of DBs below the phonon band in systems with soft-type anharmonicity \cite{Campbell2004PhysicsToday}.

The normal modes of the system are introduced via a canonical transformation
\begin{equation}
\left(\begin{array}{c}
    Q_q   \\
    P_q
\end{array}\right)
=\sum_{i=1}^N
\left(\begin{array}{c}
    x_i   \\
    p_i
\end{array}\right)
\sqrt{\frac{2}{N+1}}{\rm sin} (\frac{qi\pi}{N+1}),
\end{equation}
where $Q_q$ and $P_q$ represent the normal coordinates and their conjugate momenta, respectively.
The Hamiltonian in term of normal coordinates becomes
\begin{equation}
\begin{split}
   H =&\sum_{q=1}^N \frac{1}{2}\left[P_q^2 + \omega_q^2 Q_q^2\right]\\
    &+\sum_{q,l,m,n=1}^{N} \frac{{\Gamma}_{q,l,m,n}}{8(N+1)} C_{q,l,m,n} Q_q Q_l Q_m Q_n.
\end{split}
\end{equation}
The coefficient ${\Gamma}_{q,l,m,n}$, characterizing the strength of nonlinearity, is given by
\begin{equation}
{\Gamma}_{q,l,m,n} =  {\beta \Omega_q \Omega_l \Omega_m \Omega_n+ \phi_4},
\end{equation}
and $ C_{q,l,m,n}$ is
\begin{equation}
C_{q,l,m,n} = \sum_{\pm} \left[ \delta_{q \pm l \pm m \pm n, 0} - \delta_{q \pm l \pm m \pm n, \pm 2(N+1)}\right].
\end{equation}
The Kronecker delta functions $ \delta_{q \pm l \pm m \pm n, [0,\pm 2(N+1)]}$ define the selection rules governing the intermodal coupling. In the absence of nonlinearity the quadratic energy of mode $q$ is given by $E_q=\left(P_q^2 + \omega_q^2 Q_q^2\right)/2$.
The equation of motion for mode $q$ in the presence of nonlinearities reads
\begin{equation}
\label{EofM}
\begin{aligned}
    \ddot{Q}_q +& \omega_q^2 Q_q = -\sum_{l,m,n=1}^{N} \frac{{\Gamma}_{q,l,m,n}}{2(N+1)} C_{q,l,m,n}\,Q_l Q_m Q_n.
\end{aligned}
\end{equation}

\subsection{Numerical Method for Searching QBs}

This section briefly revisits the numerical method for obtaining QBs in nonlinear systems \cite{Flach2005PhysRevLett,Flach2006PhysRevE,Karve2024Chaos}.
According to its definition, QB is the periodic orbit in the phase space spanned by the normal coordinates, where the points are parameterized by the $2N$-dimensional vector $\bm{s} = \{Q_1, Q_2, \ldots, Q_{N}, P_1, P_2, \ldots, P_{N}\}$.
Consequently, finding QBs involves identifying periodic orbits within this phase space.
To achieve this, the Poincar{\'e} map technique is utilized.
In the weak nonlinearity regime, QB solutions can be viewed as perturbations of the normal mode orbits in the harmonic limit, representing the trivial QB solutions.

The process begins by exciting the system with the seed mode $q_0$ under the initial conditions that all sites are placed at the equilibrium positions and the initial momenta are formulated in normal coordinates as
\begin{equation}
P_q=
\begin{cases}
\sqrt{2E_{\rm total}}, &q=q_0,\\
0, &{\rm otherwise},
\end{cases}
\end{equation}
where $E_{\rm total}$ is the initial excitation energy.
This setup corresponds to the point $\bm{s}_0 = \{0, \ldots, P_{q_0}=\sqrt{2E_{\rm total}}, \ldots, 0\}$ in the phase space, lying on the Poincaré section $S: \{Q_{q_0} = 0, P_{q_0} > 0\}$. Starting from $\bm{s}_0$, the equations of motion for the whole system are numerically integrated until the trajectory returns to the section $S$ again at a new point $\bm{s}_1$. This procedure defines the Poincar{\'e} map $\bm{s}_1 = \bm{I}(\bm{s}_0)$.
The fixed points of the map $\bm{s^*}$, satisfying $\bm{s^*}= \bm{I}(\bm{s^*})$, correspond to periodic orbits and are identified as the potential QB solutions of the system.

Newton’s method is employed to search for the fixed points of the map. To resolve the degeneracy caused by energy conservation and enable the application of the implicit function theorem, the iterations are performed in a reduced phase space of dimension $2N-2$, which is parameterized by the vector
$\bm{r} = \{Q_1,Q_2, \ldots, Q_{q_0-1}, Q_{q_0+1}, \ldots, Q_{N}, P_1,P_2,$ $\ldots, P_{q_0-1}, P_{q_0+1}, \ldots, P_{N}\}$ by excluding the components $Q_{q_0}$ and $P_{q_0}$.
The fixed point of the system is determined as the root of the equation
\begin{equation}\label{mappingG}
\bm{G}(\bm{r}) = \bm{I}(\bm{r}) - \bm{r} = \bm{0}.
\end{equation}
The Newton matrix $\bm{N}$ for the vector function $\bm{G}(\bm{r})$ is defined as
\begin{equation}\label{Newtonmatrix}
\bm{N}_{ij} = \frac{\partial G_i (\bm{r})}{\partial r_j} = \frac{\partial I_i (\bm{r})}{\partial r_j} - \delta_{ij},
\end{equation}
where $\partial I_i (r)/\partial r_j$ is the Jacobian matrix of the mapping $\bm{I}$ on the Poincaré section $S$. The iteration process follows
\begin{equation}\label{iterationprocess}
\bm{r}' = \bm{r} - \bm{N}^{-1}\bm{G}(\bm{r}).
\end{equation}
After identifying the fixed point $\bm{r^*}$, we adjust the component $P_{q_0}$ to maintain energy conservation, ensuring that $P_{q_0} = \sqrt{2E - \sum_{q \neq q_0} P_q^2}$. The iteration is carried out until the convergence criterion $\left\| \bm{G} \right\| < 10^{-9}$ is satisfied, where $\left\| \bm{G} \right\|=\max \{|\bm{G}|\}$. This stringent condition ensures the precision necessary for accurately identifying the periodic orbit.

\begin{figure}
    \centering
    \includegraphics[width=1\linewidth]{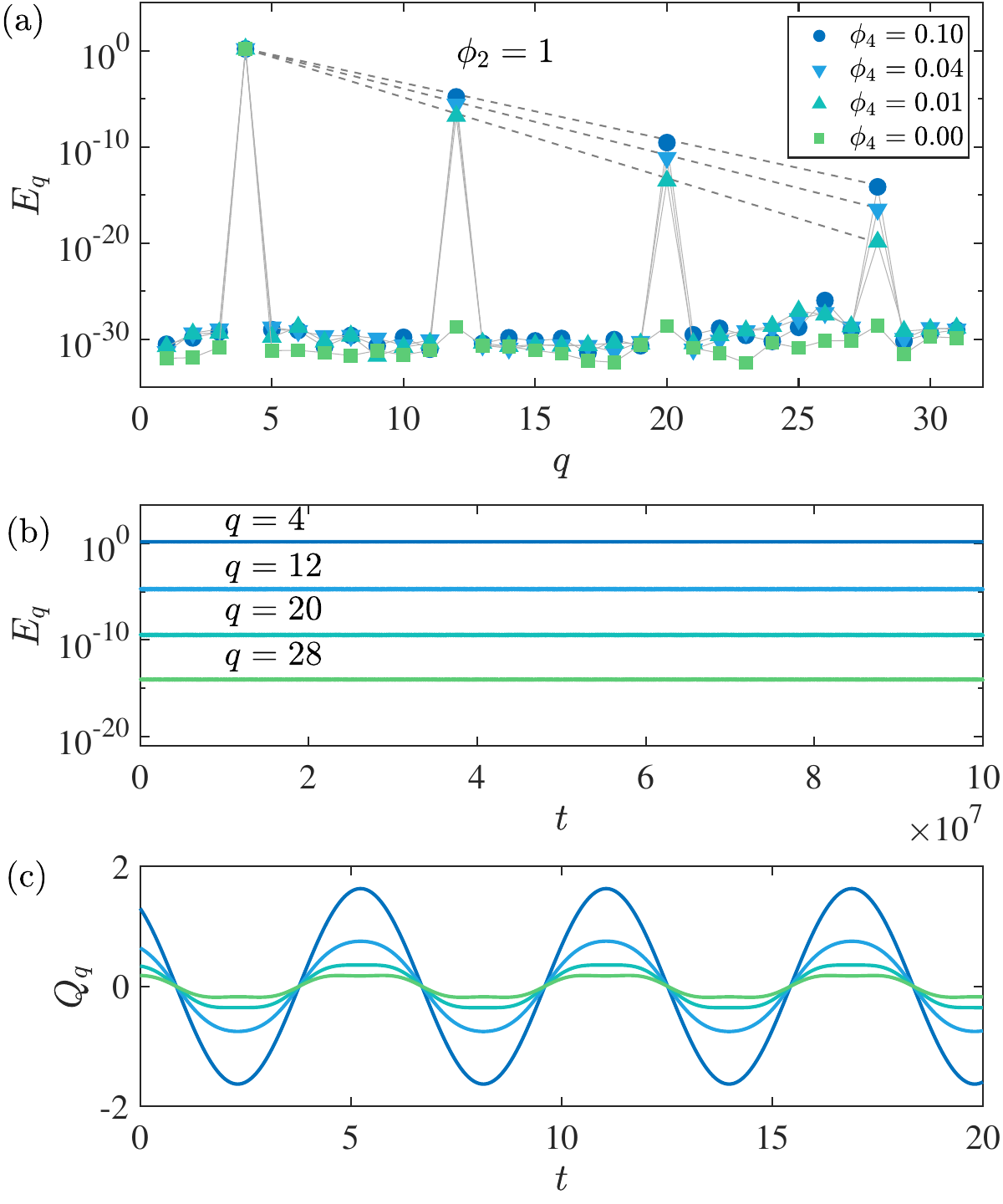}
    \caption{(a) Energy spectra of QB solutions for varying nonlinear parameters $\phi_4$, with fixed $\phi_2 = 1$, $\beta=0$, $N = 31$, seed mode $q_0 = 4$, and excitation energy $E_{\rm{total}} = 1.53$. Gray dashed lines represent exponential fits based on Eq. (\ref{eq:expon_dash}). (b) Temporal evolution of the harmonic energies of the dominant modes for a QB solution with $\phi_2 = 1$ and $\phi_4 = 0.1$. (c) Time evolution of the normal coordinates $Q_q$, starting from the final time $t = 1.0\times 10^8$ in (b). To improve visualization, the amplitudes of modes 12, 20, and 28 are scaled by the factors of $1.70 \times 10^2$, $2.25 \times 10^4$, and $3.00 \times 10^6$, respectively.
    }
    \label{Fig2_QBs}
\end{figure}

\section{QBs in Case I with $\beta=0$}\label{sec3}
\subsection{QB Solutions}

To isolate the effects of the on-site potential, we first examine the case where $\beta=0$, with the quartic on-site term being the sole source of nonlinearity. Following the outlined methodology, we calculate the QB solutions by fixing the system size at $N=$31 and setting the specific energy to $\varepsilon = 1.53 / 31$, while varying the parameters of the on-site potential, $\phi_2$ and $\phi_4$.
As a representative example, we initialize our analysis with the seed mode $q_0 = 4$, whose harmonic period is approximately $T_{q_0} = 2\pi / \omega_4 \approx 5.853$.
Numerical integration was performed with the $SBAB_2$ symplectic integrator \cite{Skokos2009PhysRevE, Skokos2010PhysRevE}, with a time step of $dt = 0.01$, ensuring precise energy conservation over the course of the extended simulations.

Figure \ref{Fig2_QBs}(a) depicts the energy spectra of QB solutions for varying $\phi_4$, with $\phi_2 = 1$ held constant. In the harmonic limit ($\phi_4 = 0$), energy remains entirely localized in the seed mode $q_0 = 4$, while other modes exhibit negligible energy, as depicted by the green squares.
Here, the system reduces to a chain of harmonic oscillators, and the energy remains indefinitely trapped in the seed mode. This serves as a trivial example of a QB solution with a fully compact energy distribution in mode space.
When $\phi_4 \neq 0$, nonlinearity introduces mode coupling, redistributing the energy initially concentrated in the seed mode among other modes. This redistribution adheres to selection rules imposed by the quartic on-site potential. The resulting energy spectra display pronounced exponential localization, well-approximated by the gray dashed lines with the expression
\begin{equation}
E(q) = E_{q_0}\exp(-\lambda q),
\label{eq:expon_dash}
\end{equation}
where $\lambda$ quantifies the localization strength. For $\phi_4 = 0.01$, 0.04, and 0.10, the values of $\lambda$ are 2.008, 1.662, and 1.435, respectively. As the nonlinearity parameter $\phi_4$ increases, more energy flows into higher-frequency modes, which results in a decrease in the parameter $\lambda$, indicating a less localized QB solution.

The time-periodic nature of QBs is another defining characteristic. Figure \ref{Fig2_QBs}(b) illustrates the energy evolution of dominant modes for $\phi_2 = 1$ and $\phi_4 = 0.1$ over a time scale of $1.0\times 10^8$, and the consistent spacing between adjacent traces on a logarithmic scale further confirms the exponential localization of QBs.
It is clear that the energies of all the modes remain practically constant throughout the simulation, signifying the stability of the QB solution. Importantly, the absence of visible energy exchange between modes suggests that QB dynamics are confined to a low-dimensional torus in phase space, which reflects the remarkable ability of QBs to sustain localized energy distributions over exceptionally long timescales, even in the presence of nonlinearity.
To further investigate this time-periodic behavior, Fig. \ref{Fig2_QBs}(c) displays the corresponding time evolutions of the normal coordinates $Q_q$ over several oscillation periods, starting from the final time $t=1.0\times 10^8$ in Fig. \ref{Fig2_QBs}(b).
Each mode exhibits regular oscillations with an identical period of $T_b = 2\pi / \hat{\omega}_b \approx 5.830$, where $\hat{\omega}_b$ denotes the characteristic frequency of QB. This consistent periodicity indicates that all modes remain synchronized, maintaining precise phase coherence throughout their evolution. This is a hallmark of QB dynamics, and highlights the intrinsic periodicity of QB solutions in nonlinear systems.

\subsection{Stability Analysis Based on Floquet Theory}

To address QB stability, we calculate the Floquet multipliers, which quantify the growth or decay of the perturbations along the periodic trajectory of QBs. As revealed in Fig. \ref{Fig2_QBs}, QB solutions feature an exponential localization in the energy spectrum, with the majority of the energy concentrated in the seed mode.
This property allows for an approximation of the QB trajectory as $Q_{q}(t)=\delta_{qq_0}A_q\cos(\omega_qt)$, with $A_q$ being the amplitude of mode $q$. By ignoring the interaction between the seed mode and other modes, the equation of motion for mode $q_0$ simplifies to
\begin{equation}\label{simE}
\ddot{Q}_{q_0} + \omega_{q_0}^2 Q_{q_0} + \frac{3 \phi_4}{2(N+1)}  Q_{q_0}^3 = 0.
\end{equation}
In the weakly nonlinear regime, where $\phi_4\varepsilon \ll 1$, an approximate solution for $Q_{q_0}(t)$ can be derived as \cite{Yoshimura2000PhysRevE}
\begin{equation}
    Q_{q_0}(t)=a \sqrt{N+1}\,\cos (\omega'_{q_0} t).
\end{equation}
Here $a^2=2 \left(\sqrt{\omega_{q_0}^4 + 6 \phi_4\varepsilon }-\omega_{q_0}^2\right)/(3 \phi_4)$, and
\begin{equation}
\omega'_{q_0} = \omega_{q_0} \left\{ 1 + \frac{9}{8} (\frac{\phi_4}{\omega_{q_0}^4}\varepsilon) - \frac{621}{256} (\frac{\phi_4}{\omega_{q_0}^4}\varepsilon)^2 \right\} + O(\varepsilon^3),
\end{equation}
is the frequency of mode $q_0$ corrected by the frequency shifts due to nonlinearity.
Notably, the frequency $\omega'_{q_0}$ aligns with the fundamental frequency $\hat{\omega}_b$ of QBs, as the latter is primarily governed by the periodic oscillation of the seed mode.

\begin{figure}
\centering
\includegraphics[width=\linewidth]{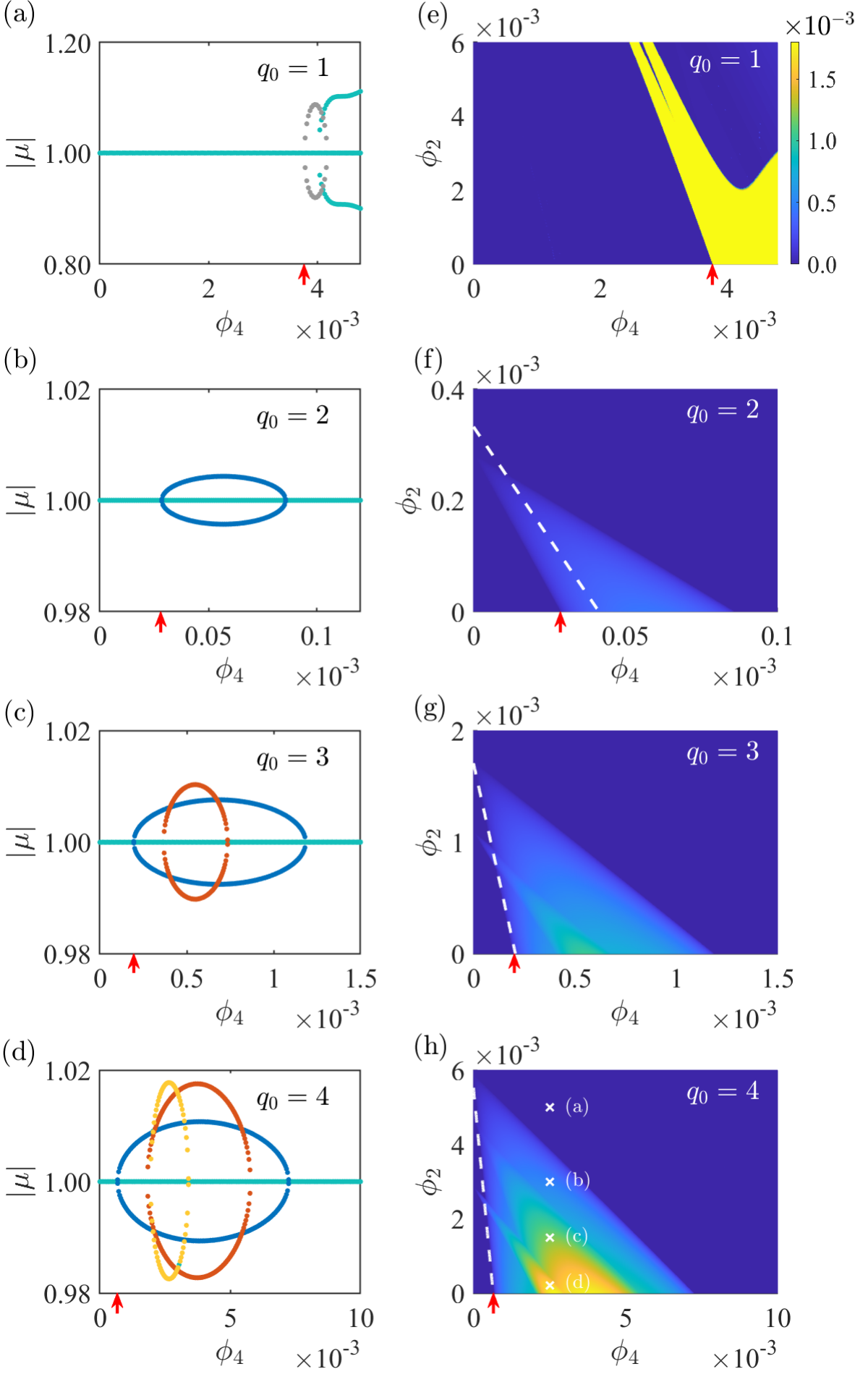}
\caption{(a)-(d) Floquet multipliers $|\mu_i|$ of QB solutions as functions of the nonlinear parameter $\phi_4$  with the excitation energy $E_{\rm total}=1.53$ and $\phi_2=0$. (e)-(h)  Contour plot of the deviation magnitude of the largest multiplier from the unit circle, i.e., $|\mu_1|-1$, in the parameter space spanned by $\phi_2$ and $\phi_4$. All panels share a common color bar. The dashed white lines represent the theoretical instability threshold given by Eq. (\ref{Eqphi2phi4}).
In all panels, the indices of the seed modes $q_0$ are indicated in the top-right corner of each panel, and the instability thresholds $\phi_{4c}$ are indicated by red arrows. }
\label{fig:localization}
\end{figure}

To evaluate the linear stability of a given QB solution, an infinitesimal perturbation $ \xi_q$ is introduced around the QB solution, i.e., $ Q_q = \hat{Q}_q + \xi_q $. The evolution of these perturbations is governed by coupled Hill equations
\begin{equation}\label{AVE}
\begin{aligned}
    \ddot{\xi}_q + \omega_q^2 \xi_q +\sum_{l,m,n=1}^{N} \frac{3\phi_4}{2(N+1)} C_{q,l,m,n} \, \hat{Q}_l \hat{Q}_m \xi_n = 0.
\end{aligned}
\end{equation}
According to the Floquet theory \cite{Nayfeh1995NonlinearOscillations,Yoshimura2000PhysRevE}, the solutions of Eq. (\ref{AVE}) at $t=0$ and $t=T_b=2\pi/\hat{\omega}_b$ are related through the monodromy matrix $\bm{M}$
\begin{equation}
\left(\begin{array}{c}
    \bm{\xi}(T_b)   \\
    \dot{\bm{\xi}} (T_b)
\end{array}\right)
=\bm{M}
\left(\begin{array}{c}
    \bm{\xi}(0)   \\
    \dot{\bm{\xi}} (0)
\end{array}\right),
\end{equation}
where $\bm{\xi}=[\xi_1,...,\xi_N]^T$ and $\dot{\bm{\xi}}=[\dot{\xi}_1,...,\dot{\xi}_N]^T$.
The Floquet multipliers $\{\mu_i \in \bm{C}, i=1,...,2N\}$ are the eigenvalues of matrix $\bm{M}$. Ordered by their moduli, i.e., $|\mu_1| \geq... \geq |\mu_{2N}|$, these multipliers exhibit symmetry properties due to the symplectic nature of the Hill equation, satisfying the relationships $\mu_{1}=\mu_{2N}^{-1}, ..., \mu_{N}=\mu_{N+1}^{-1}$. For a QB to be stable, all Floquet multipliers have to lie on the unit circle in the complex plane, with $|\mu_i|=1$ for all $i$. This condition ensures that the amplitudes of perturbations remain constant over time, maintaining the QB’s stability. Conversely, if any Floquet multiplier deviates from the unit circle, the QB becomes unstable, with the degree of deviation quantifying the instability rate.

Figure \ref{fig:localization} systematically  explores the impact of on-site potentials on the stablity of QBs. For simplicity, we first focus on the system with only quartic on-site potentials with $\phi_2=0$, and examine how the Floquet multipliers depend on the strength of quartic on-site potentials $\phi_4$, as shown in Figs. \ref{fig:localization}(a)–\ref{fig:localization}(d).
In the weak nonlinear regime, all QBs remain stable, with all the multipliers falling on the line of $\vert \mu\vert =1.0$.
However, when $\phi_4$ surpasses a critical threshold $\phi_{4c}$, marked by the red arrows in each panel, a bifurcation emerges in the Floquet spectrum. Specifically, certain eigenvalues develop moduli greater than unity, while their reciprocals fall below unity. This spectral shift signals a loss of stability, and the QB enters an unstable regime, with energy leaking from the initially localized QB modes to other modes.

In addition, Fig. \ref{fig:localization} offers deeper insights into the diverse instability mechanisms of QBs, intricately linked to the choice of seed mode $q_0$, which is demonstrated by the strikingly different patterns in the Floquet multiplier spectra.
For $q_0 = 1$, the Floquet multiplier spectrum exhibits a parabolic structure [Fig. \ref{fig:localization}(a)], reminiscent of the instability mechanisms observed in FPUT systems \cite{Flach2005PhysRevLett}.
However, for seed modes $q_0 = 2, 3, 4$, instability exhibits a more complex behavior characterized by the emergence of ring-like structures in the multiplier spectrum [Figs. \ref{fig:localization}(b)-(d)], which resemble instability islands observed in the FPUT lattices with long-range interaction, as reported in a prior study \cite{PRE2015Instabilities}.
Moreover, the complexity of the Floquet multiplier spectrum increases systematically with the seed mode $q_0$. Specifically, the number of ring structures in the spectrum is $q_0 - 1$, meaning that seed modes with higher frequency possess a greater number of distinct instability channels.

To explore the combined effect of the quadratic and quartic on-site potential coefficients on QB stability, Figs. \ref{fig:localization}(e)–(h) display the contour maps of the deviation of the largest Floquet multiplier from unity, $\vert\mu_1\vert-1$, as a function of $\phi_2$ and $\phi_4$, with well-defined boundaries demarcating instability regions.
For $q_0=1$, as $\phi_2$ increases, the instability region divides into two distinct regions, and the Floquet multiplier spectrum experiences a transition from the parabolic structure to two separate rings [Fig. \ref{fig:localization}(e)].
In contrast, for $q_0 = 2$, the instability region assumes a triangular shape [Fig. \ref{fig:localization}(f)], where QBs are unstable inside and stable outside the triangle.
Here $\phi_2$ can serve as a control parameter. When $\phi_2 = 0$, the system reaches its maximum instability threshold $\phi_{4c}$ marked by the red arrow, corresponding to the widest instability interval. As $\phi_2$ increases, $\phi_{4c}$ progressively decreases, and the instability region contracts, vanishing entirely at approximately $\phi_2 = 2.84\times 10^{-4}$, beyond which QBs remain stable for any $\phi_4$ value.
For seed modes $q_0 = 3$ and $q_0 = 4$, the instability phase diagrams exhibit more intricate structures, as shown in Figs. \ref{fig:localization}(g) and \ref{fig:localization}(h), with multiple overlapping instability regions. Despite these intricacies, the general trends observed for $q_0 = 2$ are retained, with $\phi_2$ and $\phi_4$ acting as key parameters in regulating QB stability.

\begin{figure}[t]
\centering
\includegraphics[width=\linewidth]{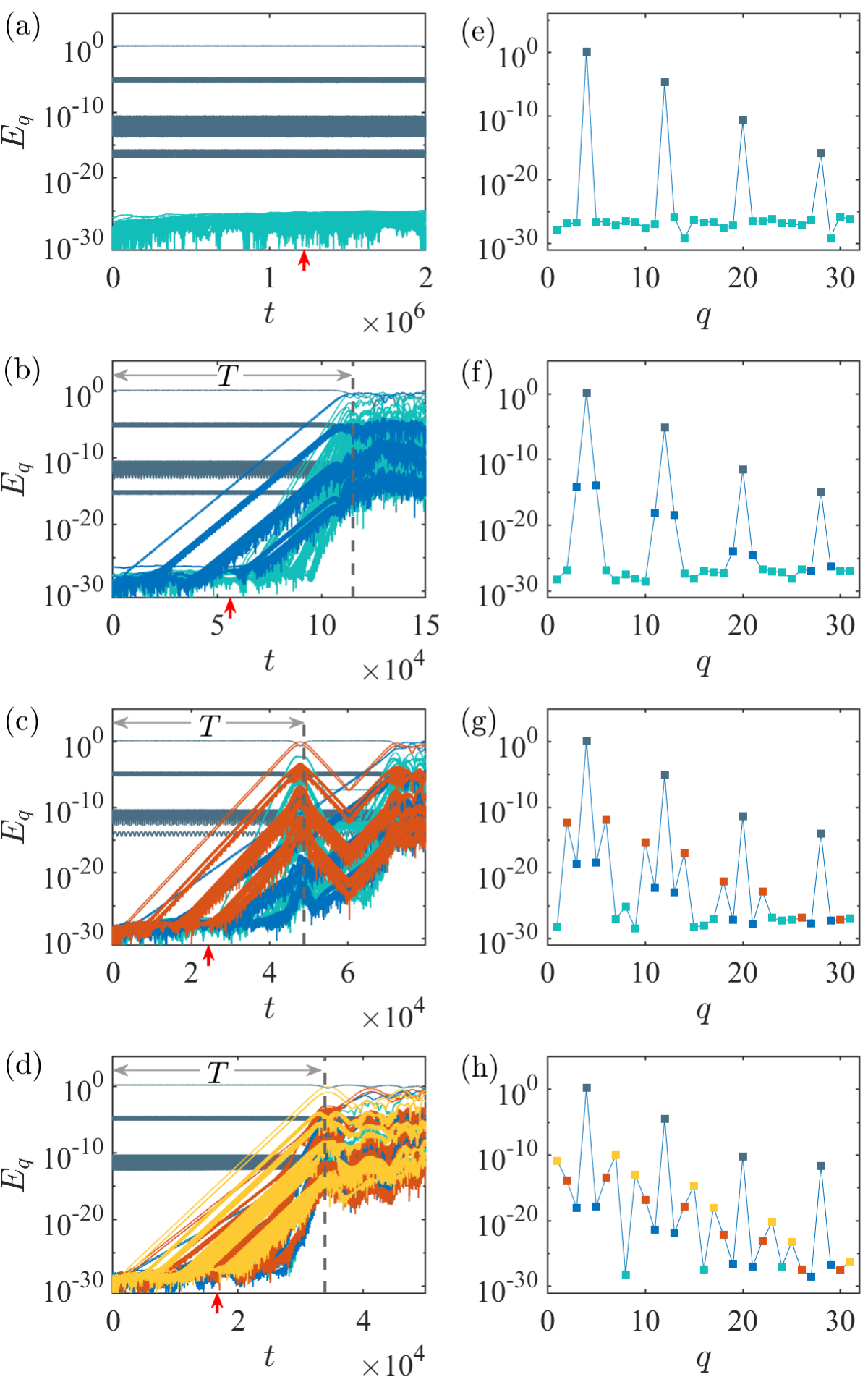}
\caption{(a)-(d) Temporal evolution of mode energies $E_q$ for systems with parameters indicated by the white crosses in Fig. \ref{fig:localization} (h), where $|\mu_1|-1$ are $4.85\times 10^{-3}$ (b), $1.19\times 10^{-2}$ (c), $1.65\times 10^{-2}$ (d), respectively. The dashed lines mark the relaxation time $T$ derived from Eq. (\ref{eq:relaxationTime}). (e)-(h) Energy spectra corresponding to panels (a)-(d), captured at the instants indicated by red arrows. In all cases, the seed mode is $q_0=4$.}
\label{fig:Dynamics}
\end{figure}

\subsection{Instability Dynamics}

To delve into the instability mechanism of QBs, we track the temporal evolution of mode energy under various system parameters.
Figures \ref{fig:Dynamics}(a)-\ref{fig:Dynamics}(d) show the simulation results for systems with the parameters marked by white crosses in Fig. \ref{fig:localization}(h).
For case (a), the multiplier $|\mu_1|$ is effectively unity, signifying a stable QB. As a result, the energy of dominant modes evolved in QBs remains unchanged throughout the evolution, as illustrated in Fig. \ref{fig:Dynamics}(a).
This stability is further confirmed by the energy spectrum at
$t\simeq 1.25\times 10^5$, where the energy is predominantly concentrated in the QB modes, with negligible contributions from other modes [Fig. \ref{fig:Dynamics}(e)].
In contrast, case (b) exhibits a deviation of $|\mu_1|$ from unity, signaling the onset of QB instability.
Figures \ref{fig:Dynamics}(b) and \ref{fig:Dynamics}(f) reveal that the nearest-neighbor modes of all modes $q$ evolved in the QB, specifically $\{q \pm 1\}$, gradually lose stability, characterized by an exponential growth in their energy, as highlighted by the blue lines in Fig. \ref{fig:Dynamics}(b). The instability triggers energy redistribution, leading to QB destabilization.
The instability becomes more pronounced in cases (c) and (d), where multiple Floquet multipliers exceed unity. This implies that additional mode pairs become unstable, leading to more complex energy transfer dynamics. In case (c), the unstable mode set expands to $\{q \pm 1, q \pm 2\}$, as shown in Fig. \ref{fig:Dynamics}(c) and the corresponding spectrum in Fig. \ref{fig:Dynamics}(g).
Similarly, for case (d), the unstable modes include $\{q \pm 1, q \pm 2, q \pm 3\}$, as indicated in Figs. \ref{fig:Dynamics}(d) and \ref{fig:Dynamics}(h).

Based on these observations, we hypothesize that the instability of QBs is primarily governed by the parametric resonance of modes $k = q_0 - m$ and $ l= q_0 + m$, where $m=1,2,...,q_0-1$. These resonances are primarily driven by the seed mode $q_0$. For each mode pair, the equations governing the perturbations, given in Eq. (\ref{AVE}), can be reformulated as
\begin{equation}
\begin{aligned}
&\frac{d^2 \xi_k}{dt^2} + \omega_k^2 \xi_k + \frac{3\phi_4 \hat{Q}_{q_0}^2}{2(N+1)} \left[2\xi_k + \xi_l \right]=0, \\
&\frac{d^2 \xi_l}{dt^2} + \omega_l^2 \xi_l + \frac{3\phi_4 \hat{Q}_{q_0}^2}{2(N+1)} \left[2\xi_l + \xi_k \right]=0,
\end{aligned}
\label{eq:couplMat}
\end{equation}
where $\hat{Q}_{q_0}(t) = a \sqrt{N+1}\,\cos (\hat{\omega}_b t)$ represents the oscillatory behavior of the seed mode $q_0$.
To proceed, the dimensionless time variable $ \tau = \omega_{q_0} t$ is introduced, allowing Eq. (\ref{eq:couplMat}) to be recast in the following form
\begin{equation}\label{CoupledMathieu}
\begin{aligned}
    & \frac{d^2\xi_k}{d\tau^2}  + r_k^2 \xi_k
    =-\gamma \left[ 1 + \cos(2\Omega \tau) \right] \left(2B_{kk} r_k^2 \xi_k +B_{kl} r_k r_l \xi_l \right), \\
    & \frac{d^2 \xi_l}{d\tau^2}  +r_l^2 \xi_l
    =-\gamma \left[ 1 + \cos(2\Omega \tau) \right] \left(B_{lk} r_l r_k \xi_k +2 B_{ll}r_l^2 \xi_l\right),
\end{aligned}
\end{equation}
where $ r_k = \omega_k/\omega_{q_0}$, $ r_l = \omega_l/\omega_{q_0} $,  $\gamma=3\phi_4a^2/(4 \omega_{q_0}^2)$, $ \Omega = \hat{\omega}_b/\omega_{q_0}$, and $B_{ij}=\omega_{q_0}^2/\omega_{i}\omega_{j}$. This formulation can be analyzed using an averaging method to delineate the instability boundaries in the $(\phi_2, \phi_4)$ parameter space  \cite{Yoshimura2000PhysRevE,Bogolyubov1961AsymptoticMI}. Specifically, the boundary separating stable and unstable regions is explicitly expressed as (see Appendix \ref{appXB} for more details)
\begin{equation}\label{Eqphi2phi4}
    \phi_2 = \sec \left(\frac{\pi  q_0}{N+1}\right) \left(1-\frac{3 \varepsilon  }{8 \zeta^2}\phi_4\right)+\cos \left(\frac{\pi  q_0}{N+1}\right)-2,
\end{equation}
where $\zeta= m \pi/[2(N+1)]$.
The theoretical predictions from Eq. (\ref{Eqphi2phi4}) are depicted by the white dashed lines in Fig. \ref{fig:localization}, which show noticeable deviations from the instability boundaries derived from Floquet analysis for $q_0=1$ and $q_0=2$.
However, for $q_0 \geq 3$, the theoretical curves closely match the instability boundaries and intersect the $x$-axis at the point marked by the red arrows, demonstrating excellent agreement (see Appendix \ref{sec:Mode56} for more cases).

\begin{figure}[t]
\centering
\includegraphics[width=\linewidth]{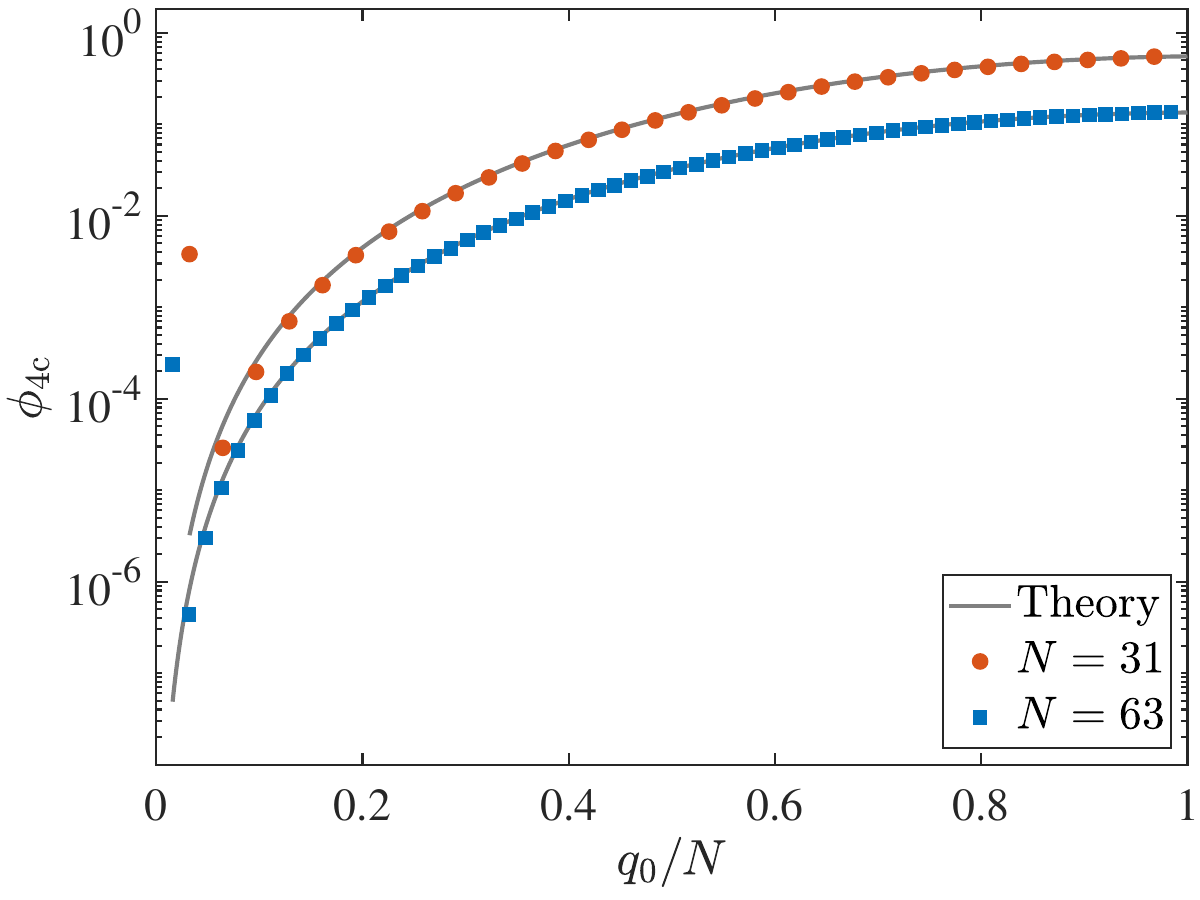}
\caption{Dependence of the stability threshold $\phi_{4c}$ on the seed mode $q_0$ for $\phi_2 =0$. The gray solid lines are the theoretical results  based on  Eq. (\ref{Eqphi4c}). The data corresponding to $q_0=N$ are not shown, as they lie outside the plotted range of the coordinate axes.}
\label{fig:Eq0}
\end{figure}

In the special case where $\phi_2 =0$, the critical value $\phi_{4c}$ for the QB instability is given by
\begin{equation}\label{Eqphi4c}
\phi_{4c} = \frac{8 \pi^2 }{3(N+1)^2 \varepsilon } \sin ^4\left[\frac{q_0 \pi}{2 (N+1)}\right],
\end{equation}
which establishes a direct relationship between the instability threshold $\phi_{4c}$, system size $N$, and seed mode $q_0$.
The critical value $\phi_{4c}$ is strongly influenced by the seed mode $q_0$, with a pronounced increase in $\phi_{4c}$ as $q_0$ increases, as illustrated in Fig. \ref{fig:Eq0}.
This trend is fundamentally different from that observed in FPUT-$\beta$ systems, where the instability threshold is independent of $q_0$ \cite{Flach2005PhysRevLett,Flach2006PhysRevE}. Such a distinct contrast highlights the critical role played by on-site potentials in modulating the stability of QBs.
Note that, these analyses are based on the assumption that the instability of QBs is primarily dominated by the parametric resonance of mode pairs $q_0 \pm m$. However, this assumption ceases to hold for specific modes located near the band edges. For $q_0 = 1$ or $q_0 = N$, the boundary constraints prevent the formation of mode pairs, thereby precluding the parametric resonance mechanism.  Consequently, numerical results for these edge modes exhibit significant deviations from the theoretical predictions based on parametric resonance.

To quantify the instability, we estimate the relaxation timescale of QBs in the unstable region. Specifically, we monitor the energy in modes other than those initially excited by the QB, defined as
\begin{equation}
\Delta E(t)=\sum_{q\notin \rm QB} \frac{1}{2}[P_q^2(t)+\omega_q^2 Q_q^2(t)].
\end{equation}
In the unstable region, $\Delta E(t)$ grows exponentially over time, given by
\begin{equation}
\Delta E(t)\approx \Delta E(0) e^{2\lambda_1 t},
\end{equation}
where the exponent $\lambda_1$ is determined from the largest characteristic exponent
$\lambda_1=\text{ln} |\mu_1|/T_{b}$.
The relaxation time $T$ is defined as the time at which $\Delta E(t)$ becomes comparable to the total energy $E_{\rm total}$ of the system, i.e., $\Delta E(T) = O(E_{\rm total})$. Using this condition, the relaxation time can be approximated as
\begin{equation}
\label{eq:relaxationTime}
T \approx \frac{1}{2\, \lambda_1} \text{ln} \frac{E_{\rm total}}{\Delta E_0},
\end{equation}
which is marked by the gray dashed lines in
Figs. \ref{fig:Dynamics}(b)-(d).
At this point, the seed mode has dissipated the majority of its energy, transferring it to other initially inactive modes. This redistribution signifies a gradual evolution toward thermal equilibrium.

\section{QBs in Case II and III}\label{sec:betaOnsite}

In this section, we extend our analysis to case II and case III, which provides a comprehensive understanding of how the combined effects of on-site potentials and nonlinear inter-site interactions influence the QB stability.

\begin{figure}[h]
\centering
\includegraphics[width=\linewidth]{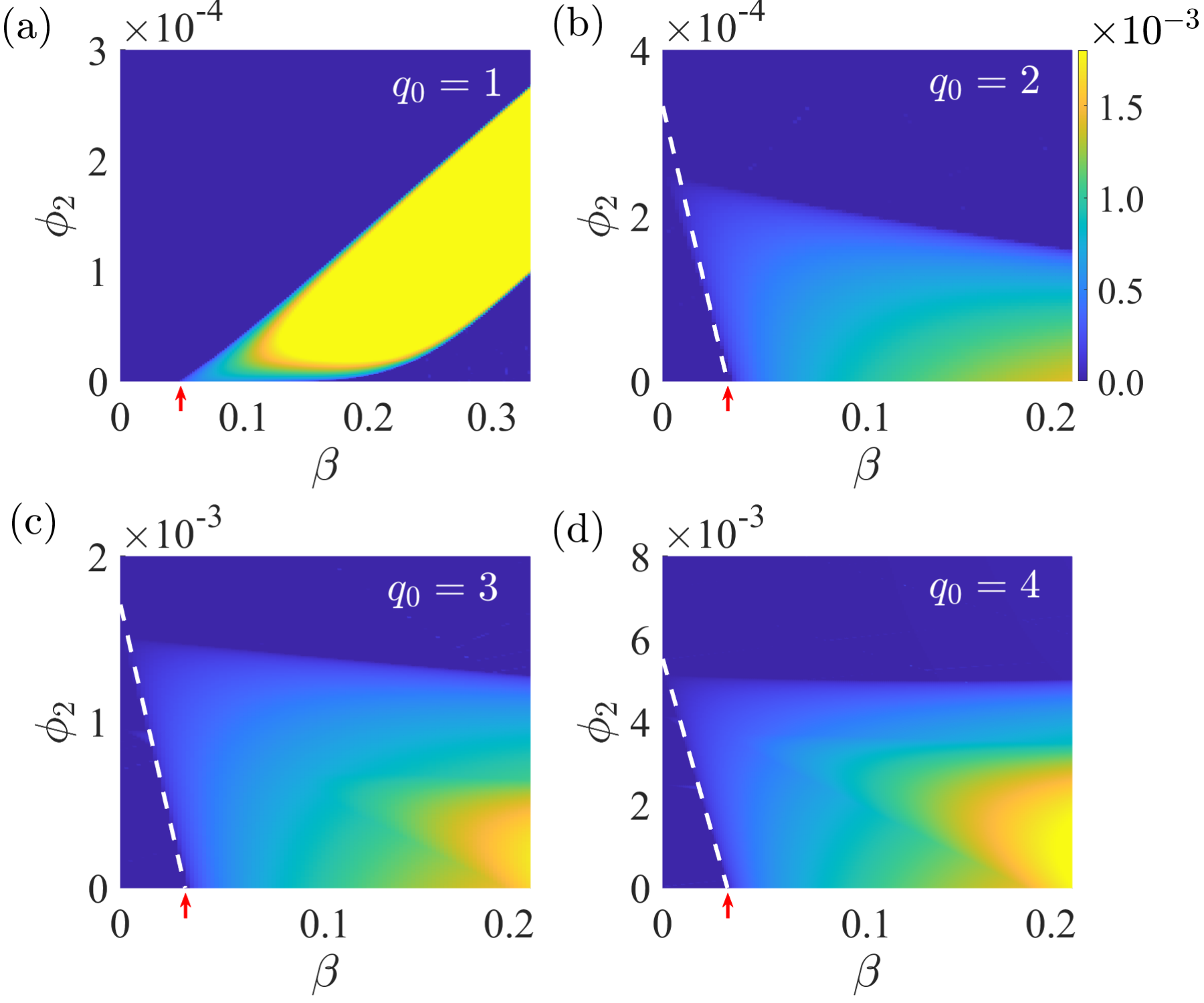}
\caption{Contour plot of the deviation magnitude of the largest multiplier from the unit circle, i.e., $|\mu_1|-1$, for case II, displayed in the plane of $\beta$ and $\phi_2$ for different seed modes: $q_0=$1(a), $q_0=$2(b), $q_0=$3(c), and $q_0=$4(d). All panels use a consistent color scale. The dashed white lines are the theoretical results based on Eq. (\ref{Eqphi2phi4}).}
\label{fig:beta2}
\end{figure}

For case II with $\phi_4=0$, the Hamiltonian contains a sum of the FPUT-$\beta$ potential and the quadratic on-site term, where the latter modifies the spectrum of normal mode frequencies.
Figure \ref{fig:beta2} illustrates the phase diagrams depicting the instability regions of QBs, where panels (a)-(d) correspond to QBs seeded by modes $q_0= 1,2,3,4$, respectively. In the linear regime ($\beta = 0$), the system reduces to a harmonic chain, where all modes represent trivial QBs without coupling, regardless of the on-site potential strength $\phi_2$.
When $\phi_2=0$, the model simplifies to the FPUT-$\beta$ system, where the instability threshold is given by $\beta_c = \pi^2 / [6 (N + 1)^2 \varepsilon]\simeq 0.034$, independent of the seed mode for $q_0\ge 2$, as indicated by the red arrows in each panel \cite{Flach2005PhysRevLett,Flach2006PhysRevE}.

\begin{figure}[h]
\centering
\includegraphics[width=\linewidth]{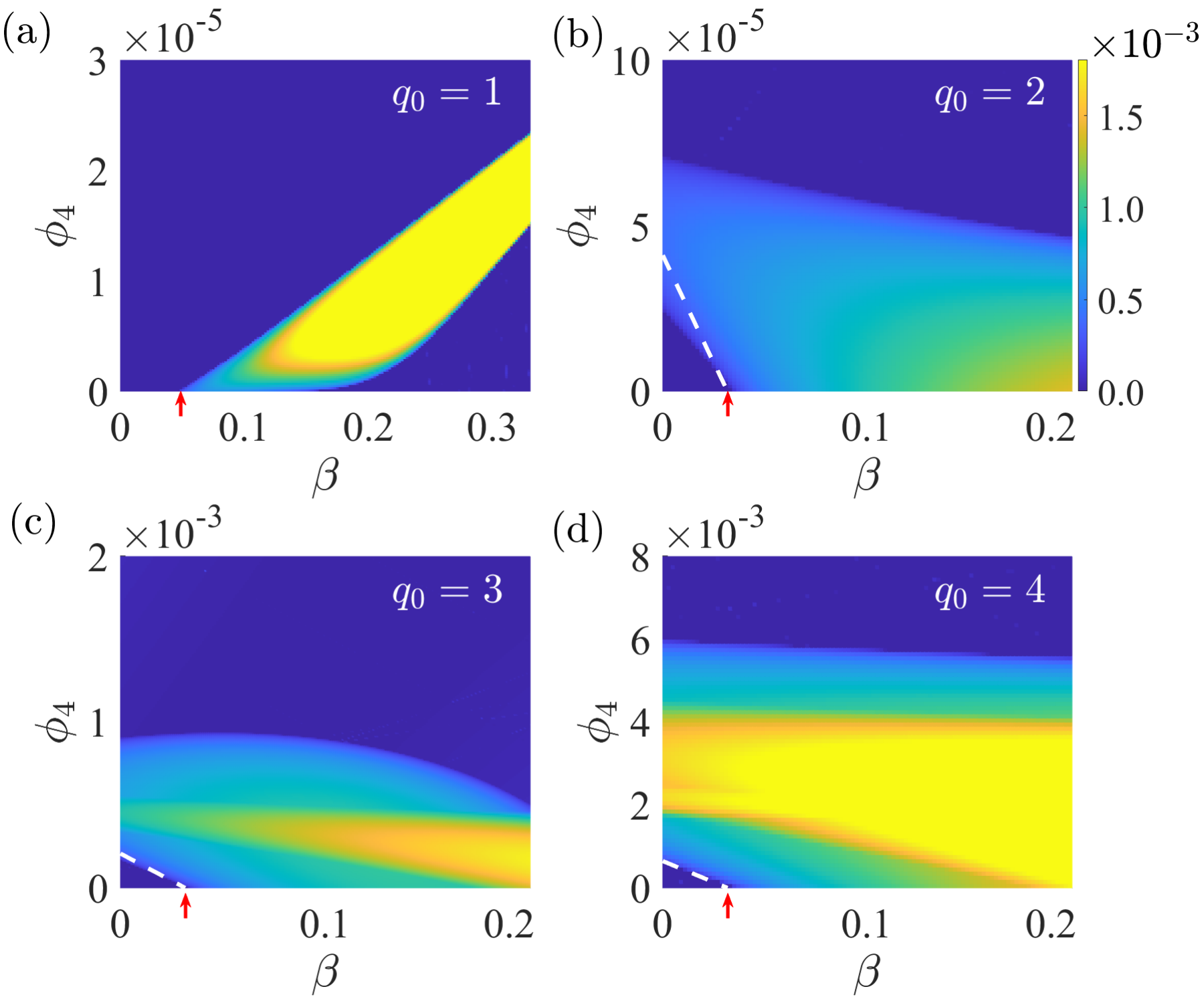}
\caption{Similar to Fig. \ref{fig:beta2}, but for case III.}
\label{fig:beta4}
\end{figure}

Analogous to case I, the instability threshold depends on the strength of the on-site potential.
For $q_0 = 1$ [Fig. \ref{fig:beta2}(a)], increasing $\phi_2$ shifts the instability threshold $\beta_c$ to higher values. It is accompanied by larger Floquet multipliers, indicating an enhanced instability rate. Conversely, for higher seed modes i.e., $q_0 = 2, 3, 4$ in Fig. \ref{fig:beta2}(b)-(d), increasing of $\phi_2$ progressively reduces the instability threshold $\beta_c$.
The transition curve separating stable and unstable regions in the $(\phi_2, \beta)$ parameter plane is derived as  (see Appendix \ref{appXB} for details)
\begin{equation}\label{EqPhi2beta}
\phi_2 = \frac{2 \left(2 \zeta^2-3 \beta  \varepsilon \right)  \sec \left(\frac{\pi  q_0}{N+1}\right)}{\zeta^2}\sin ^4\left[\frac{\pi  q_0}{2 (N+1)}\right].
\end{equation}
This expression is plotted with the white dashed lines in Figs. \ref{fig:beta2}(b)-\ref{fig:beta2}(d), which demonstrate the dependence of instability threshold on the seed mode $q_0$ and align well with numerical results for $q_0 \geq 3$.
An additional feature of the phase diagrams is that QBs remain stable when $\phi_2$ exceeds a critical value, regardless of the nonlinearity parameter $\beta$. This highlights the stabilizing role of strong on-site potentials in suppressing QB instability.

Similarly, Fig. \ref{fig:beta4} shows the phase diagrams of the instability regions for case III, which exhibit similar structures to those of case II. A key distinction arises at $\beta = 0$, where the system reduces to the one for case I with $\phi_2 = 0$. In this case, QBs are unstable within a specific parameter range of $\phi_4$.
The transition curve for instability threshold is
\begin{equation}\label{EqPhi4beta}
\phi_4 = \frac{16 \left(2\zeta^2-3 \beta  \varepsilon \right) }{3 \varepsilon }\sin ^4\left[\frac{\pi  q_0}{2(N+1)}\right],
\end{equation}
which is represented by the white dashed lines in Figs. \ref{fig:beta4}(b)-\ref{fig:beta4}(d).

\section{Conclusion and Discussion}\label{sec:Conclusion}

In conclusion, the effects of on-site potentials on the stability of QBs has been systematically investigated in nonlinear lattices, where the on-site potentials are incorporated in the Hamiltonian of the FPUT-$\beta$ system.
Using numerical simulations and analytical approaches based on Floquet theory, we identified the conditions and parameter ranges governing QB stability. A key finding is that the stability of QBs is highly sensitive to the strength of the on-site potential, where tuning the on-site potential can shift the instability threshold profoundly. In particular, strong quadratic on-site potentials are shown to stabilize QBs even in the presence of large nonlinearities, suggesting that such potentials can play a significant role in controlling QB dynamics. Our analysis demonstrates that QB instability is often driven by parametric resonance, effectively modeled by coupled Mathieu equations. The derived theoretical instability thresholds exhibit excellent agreement with numerical results, particularly for QBs originated from higher-frequency seed modes. Moreover, the sensitivity of the instability threshold on the seed mode choice highlights the intricate dependence of properties of localized excitations on the on-site potential, which is in contrast to previous studies in systems without substrate potentials.

Our study provides a comprehensive framework for understanding and analyzing the stability of QBs in nonlinear lattice systems. It highlights the importance of on-site potentials in shaping the behavior of localized excitations, thereby significantly advancing the theoretical understanding of QB dynamics.
The implications of these findings extend beyond the systems studied here. The ability to manipulate QB stability using on-site potentials opens up new possibilities for controlling localized excitations in a wide range of nonlinear analogous FPUT systems. This is especially important for systems that require localization, including the design of cold atom arrays \cite{Singh2012PhysRevA, He2018PhysRevA}, Josephson junction arrays \cite{Insulating2022prb,Dynamical2023prb}, optical waveguide arrays \cite{Dynamics1999prl,Gyrating2021pra}, among others. Furthermore, while this study focuses on one-dimensional systems, the insights gained here provide a solid foundation for future research into higher-dimensional systems, where more complex interactions and nonlinear effects may come into play.

\begin{acknowledgments}
This work was supported by National Key R\&D Program of China under Grant No. 2023YFA1407101, NSFC under Grant Nos. 11905087, 12175090, 11775101, 12247101 and 12465010, the Fundamental Research Funds for the Central Universities (Grant No. lzujbky-2024-jdzx06), NSF of Gansu Province under Grant No. 22JR5RA389, and the ‘111 Center’ under Grant No. B20063. W. Fu also acknowledges support by the Youth Talent (Team) Project of Gansu Province; the Innovation Fund from Department of Education of Gansu Province (Grant No.~2023A-106). S. Flach was supported by the Institute for Basic Science through Project Code (No. IBS-R024-D1).
\end{acknowledgments}

\appendix
\section{Derivation of the Transition Curve}\label{appXB}

This section briefly reviews the method for estimating the exponential growth rate of the solution to Eq. (\ref{CoupledMathieu}) \cite{Yoshimura2000PhysRevE}.
Infinitesimal perturbations, $ \xi_q$, are introduced around the QB solution $ \hat{Q}_q(t) $, such that $ Q_q = \hat{Q}_q + \xi_q $. The evolution of these perturbations follows the coupled Hill equations
\begin{equation}\label{AVE0}
\begin{aligned}
    \ddot{\xi}_q &+ \omega_q^2 \xi_q =
    - \sum_{l,m,n=1}^{N} \frac{3{\Gamma}_{q,l,m,n}}{2(N+1)} \, C_{q,l,m,n} \, \hat{Q}_l \hat{Q}_m \xi_n.
\end{aligned}
\end{equation}
For each mode pair, Eq. (\ref{AVE0}) can be reformulated as
\begin{equation}
\begin{aligned}
&\frac{d^2 \xi_1}{dt^2}  + \omega_1^2 \xi_1 \\
=&- \frac{3 \hat{Q}_{q_0}^2}{2(N+1)} \left[ 2{\Gamma}_{1,q_0,q_0,1}\xi_1 + {\Gamma}_{1,q_0,q_0,2}\xi_2 \right], \\
&\frac{d^2 \xi_2}{dt^2} + \omega_2^2 \xi_2 \\
=&- \frac{3 \hat{Q}_{q_0}^2}{2(N+1)} \left[2{\Gamma}_{2,q_0,q_0,2}\xi_2 + {\Gamma}_{2,q_0,q_0,1}\xi_1 \right],
\end{aligned}
\end{equation}
where $\hat{Q}_{q_0}(t) = a \sqrt{N+1}\,\cos (\hat{\omega}_b t) $. By introducing the dimensionless time variable $ \tau = \omega_{q_0} t$, this equation becomes
\begin{equation}\label{CoupledMathieu0}
\begin{aligned}
    & \frac{d^2\xi_1}{d\tau^2}  + r_1^2 \xi_1
    =-\gamma \left[ 1 + \cos(2\Omega \tau) \right] \left(2 B_{11} r_1^2 \xi_1 +B_{12} r_1 r_2 \xi_2 \right) , \\
    & \frac{d^2 \xi_2}{d\tau^2}  +r_2^2 \xi_2
    =- \gamma \left[ 1 + \cos(2\Omega \tau) \right] \left(2 B_{22} r_2^2 \xi_2 + B_{21} r_2 r_1 \xi_1 \right) ,
\end{aligned}
\end{equation}
where $ r_1 = \omega_k/\omega_{q_0}$, $ r_2 = \omega_l/\omega_{q_0} $,  $\gamma=3 {\Gamma}_{0} a^2/(4\omega_{q_0}^2 )$, $ \Omega = \hat{\omega}_b/\omega_{q_0}$, and $B_{ij}=({\Gamma}_{i,q_0,q_0,j}\omega_{q_0}^2 )/({\Gamma}_{0}\omega_i\omega_j)$.

We assume a solution to Eq. (\ref{CoupledMathieu0}) in the form
\begin{equation}\label{Solution2M}
\begin{aligned}
    \xi_i&=u_i(\tau) \text{sin} (\Omega \tau)+ v_i(\tau) \text{cos} (\Omega \tau) \\
    \frac{d\xi_i}{d\tau}&=\Omega u_i(\tau) \text{cos} (\Omega \tau)-\Omega v_i(\tau) \text{sin} (\Omega \tau)
\end{aligned}
\end{equation}
where $i=1,2$.
Then Eqs. (\ref{CoupledMathieu0}) can be rewritten as
\begin{widetext}
\begin{equation}\label{Eq_uv}
\begin{aligned}
    \frac{d}{d\tau} u_1(\tau) &= \frac{\gamma}{\Omega} \left[ \frac{1}{2} a_{13} v_1(\tau) - \frac{3}{4}  b_{12} v_2(\tau) \right. \\
    &\quad + \left( \frac{1}{2} a_{12} u_1(\tau) - \frac{1}{2}b_{12} u_2(\tau) \right) \sin(2 \Omega \tau) + \left( \frac{1}{2} a_{14} v_1(\tau) - b_{12} v_2(\tau) \right) \cos(2 \Omega \tau) \\
    &\quad + \left( -\frac{1}{2}b_{11} u_1(\tau) - \frac{1}{4}b_{12} u_2(\tau) \right) \sin(4 \Omega \tau) + \left( -\frac{1}{2}b_{11} v_1(\tau) - \frac{1}{4}b_{12} v_2(\tau) \right) \cos(4 \Omega \tau) \Bigg],\\
    \frac{d}{d\tau} v_1(\tau) &= \frac{\gamma}{\Omega} \left[ -\frac{1}{2} a_{11} u_1(\tau) + \frac{1}{4}b_{12} u_2(\tau) \right. \\
    &\quad + \left( -\frac{1}{2} a_{12} v_1(\tau) + \frac{1}{2}b_{12} v_2(\tau) \right) \sin(2 \Omega \tau) +\frac{1}{2} a_{10} u_1(\tau) \cos(2 \Omega \tau) \\
    &\quad + \left( \frac{1}{2}b_{11} v_1(\tau) + \frac{1}{4} b_{12} v_2(\tau) \right) \sin(4 \Omega \tau) + \left( -\frac{1}{2}b_{11} u_1(\tau) - \frac{1}{4}b_{12} u_2(\tau) \right) \cos(4 \Omega \tau) \Bigg],\\
    \frac{d}{d\tau} u_2(\tau) &= \frac{\gamma}{\Omega} \left[ \frac{1}{2} a_{23} v_2(\tau) - \frac{3}{4} b_{12} v_1(\tau) \right. \\
    &\quad + \left( \frac{1}{2} a_{22} u_2(\tau) - \frac{1}{2} b_{12} u_1(\tau) \right) \sin(2 \Omega \tau) + \left( \frac{1}{2} a_{24} v_2(\tau) - b_{12} v_1(\tau) \right) \cos(2 \Omega \tau) \\
    &\quad + \left(- \frac{1}{2}b_{22} u_2(\tau) -\frac{1}{4} b_{12} u_1(\tau)\right) \sin(4 \Omega \tau) + \left( -\frac{1}{4}b_{12} v_1(\tau) - \frac{1}{2}b_{22} v_2(\tau) \right) \cos(4 \Omega \tau) \Bigg],\\
    \frac{d}{d\tau} v_2(\tau) &= \frac{\gamma}{\Omega} \left[ -\frac{1}{2} a_{21} u_2(\tau) + \frac{1}{4}b_{12} u_1(\tau) \right. \\
    &\quad + \left(-\frac{1}{2} a_{22} v_2(\tau) +\frac{1}{2}b_{12} v_1(\tau)\right) \sin(2 \Omega \tau) + \frac{1}{2} a_{20} u_2(\tau) \cos(2 \Omega \tau) \\
    &\quad + \left( \frac{1}{4}b_{12} v_1(\tau) + \frac{1}{2}b_{22} v_2(\tau) \right) \sin(4 \Omega \tau) + \left(- \frac{1}{2}b_{22} u_2(\tau) -\frac{1}{4}b_{12} u_1(\tau)\right) \cos(4 \Omega \tau) \Bigg],
\end{aligned}
\end{equation}
\end{widetext}
where the coefficients $a_{ij}$ and $b_{ij}$ are given by $a_{ij}=\frac{1}{\gamma} [\Omega^2 - (1 + j \gamma B_{ii} ) r_i^2]$ and $b_{ij}=B_{ij} r_i r_j$, respectively.
Assuming $m/N << q_0/N$, this condition ensures that $r_1 \approx 1$ and $r_2 \approx 1$. In addition, for sufficiently small $\varepsilon$, one has $\Omega \approx 1$. Under these conditions, it follows that $a_{ij} \ll O (\gamma^{-1})$ and $b_{ij}=O(1)$.

The derived equation is in the standard form for the averaging method, which is applicable when $m/N \ll q_0/N$ and $\varepsilon << 1$. By calculating the second-order averaged equations, we arrive at
\begin{equation}\label{EQeigValue}
\frac{d}{d\tau} \begin{pmatrix} u_1 \\ v_1 \\ u_2 \\ v_2 \end{pmatrix} = \frac{\gamma}{4 \Omega} \begin{pmatrix} 0 & 2 \Delta_{1u} & 0 & -3 R_1 \\ -2 \Delta_{1v} & 0 & R_2 & 0 \\ 0 & -3 R_1 & 0 & 2 \Delta_{2u} \\ R_2 & 0 & -2 \Delta_{2v} & 0 \end{pmatrix} \begin{pmatrix} u_1 \\ v_1 \\ u_2 \\ v_2 \end{pmatrix},
\end{equation}
where
\begin{eqnarray}
\nonumber
\Delta_{1u} &=& a_{13} + \frac{\gamma}{32 \Omega^2} \left( 8 a_{12} a_{14} + 4 b_{11}^2 + 17 b_{12}^2 \right),\\
\nonumber
\Delta_{1v} &=& a_{11} + \frac{\gamma}{32 \Omega^2} \left( 8 a_{10} a_{12} + 4 b_{11}^2 + b_{12}^2 \right),\\
\nonumber
\Delta_{2u} &=& a_{23} + \frac{\gamma}{32 \Omega^2} \left( 8 a_{22} a_{24} + 4 b_{22}^2 + 17 b_{12}^2 \right),\\
\nonumber
\Delta_{2v} &=& a_{21} + \frac{\gamma}{32 \Omega^2} \left( 8 a_{20} a_{22} + 4 b_{22}^2 + b_{12}^2 \right),
\end{eqnarray}
and
\begin{eqnarray}
\nonumber
R_1 &=& b_{12} + \frac{\gamma}{48 \Omega^2}
\left[4 b_{12} (2 a_{12} + a_{14} + a_{24} + 2 a_{22}) \right .\\
\nonumber
 && \left . - 2 b_{12} (b_{11} + b_{22}) \right],\\
\nonumber
 R_2 &=& b_{12} + \frac{\gamma}{16 \Omega^2} \left[ 4 b_{12} (a_{10} + a_{20}) - 2 b_{12} (b_{11} + b_{22}) \right].
\end{eqnarray}
The eigenvalues $\lambda$ of the coefficient matrix in Eq. (\ref{EQeigValue}) can be explicitly obtained as
\begin{equation}
    \lambda = \pm \frac{\gamma}{4 \Omega} \sqrt{G \pm 2 \sqrt{F}},
\end{equation}
where
\begin{equation}
\begin{aligned}
    F =& (\Delta_{1u} R_2 + 3 \Delta_{2v} R_1)(\Delta_{2u} R_2 + 3 \Delta_{1v} R_1)\\
      +& (\Delta_{1u} \Delta_{1v} - \Delta_{2u} \Delta_{2v})^2,
\end{aligned}
\end{equation}
and
\begin{equation}
G = -3 R_1 R_2 - 2 (\Delta_{1u} \Delta_{1v} + \Delta_{2u} \Delta_{2v}).
\end{equation}

The polynomials $F$ and $G$ are functions of $\varepsilon$. By examining their dependence on $\varepsilon$, we observe that, for fixed values of $m/N$ and $q_0/N$, the polynomial $F$ experiences a transition from positive to negative values as $\varepsilon$ increases, while $G$ remains negative throughout. Moreover, the magnitude of $|G|$ is much larger than $|F|$.
When $F$ is positive, all the eigenvalues $\lambda$ are purely imaginary, implying that the solution to Eq. (\ref{EQeigValue}) is stable. In this scenario, the system does not exhibit any exponential growth, and the solution remains bounded.
Conversely, when $F$ becomes negative, the eigenvalues take the form $\pm(x\pm iy)$, with $x$ and $y$ being real numbers. In this case, the solution of Eq. (\ref{EQeigValue}) becomes unstable, and the system undergoes exponential growth at a rate determined by the positive real part of the eigenvalues $\operatorname{Re}[\lambda]$.
This exponential growth rate is also applicable to the solution $(\xi_1,\xi_2)$ of the coupled Mathieu equations, i.e., Eq. (\ref{CoupledMathieu}), as the solution's expressions are linear in $u_i$ and $v_i$.

When $F < 0$ and $G < 0$, one of the eigenvalues can be written as
$$\lambda= \frac{\gamma}{4 \Omega} \sqrt{G + 2 \text{i} \sqrt{|F|}} = \frac{\gamma}{4 \Omega} \left(G^2 + 4 |F|\right)^{1/4} e^{\text{i} \theta / 2},$$
where $\cos \theta = G/\sqrt{G^2 + 4 |F|}$ and
$\cos\left(\theta/2\right) =  \sqrt{(1 + \cos \theta)/2}$. In this case, the solution of Eq. (\ref{EQeigValue}) is unstable and grows at a rate given by
\begin{equation}
    \operatorname{Re}[\lambda] = \frac{\gamma}{4 \sqrt{2} \Omega} \left[G + \sqrt{G^2 + 4 |F|} \right]^{1/2}.
\end{equation}
For the regime where $-F \ll 1$, the growth rate $\operatorname{Re}[\lambda]$ can be expanded as
\begin{equation}\label{Relambda}
    \operatorname{Re}[\lambda] \approx \frac{\gamma}{4 \sqrt{2} \Omega} \left[G - G \left(1 + \frac{2 |F|}{G^2}\right)\right]^{1/2} = \frac{\gamma}{4 \Omega} \sqrt{\frac{F}{G}}.
\end{equation}
By introducing $\zeta = \pi m/[2(N+1)]$ and expanding Eq. (\ref{Relambda}) to the order of $\varepsilon\zeta^4$, we arrive at
\begin{widetext}
\begin{equation}\label{RElambda}
\begin{aligned}
    \operatorname{Re}[\lambda]\approx &\frac{\zeta\, \csc^4\left(\frac{\pi q_0}{2 (N + 1)}\right)}{\left[4 + \phi_2 \, \csc^2\left(\frac{\pi q_0}{2 (N + 1)}\right)\right]^2} \sqrt{3 - 2 (2 + \phi_2) \cos\left(\frac{\pi q_0}{N + 1}\right) + \cos\left(\frac{2 \pi q_0}{N + 1}\right)} \\
    \times&\left[\frac{3}{4} \varepsilon (6 \beta + \phi_4) + 2 \left[\zeta^2 (2 + \phi_2) - 3 \beta \varepsilon\right] \cos\left(\frac{\pi q_0}{N + 1}\right) - 3\zeta^2 - (\zeta^2 - \frac{3}{2} \beta \varepsilon) \cos\left(\frac{2 \pi q_0}{N + 1}\right)\right]^{1/2}.
\end{aligned}
\end{equation}
\end{widetext}

Based on Eq. (\ref{RElambda}), the transition curve for the QB instability can be derived. In the special case where $\phi_2=0$ and $\phi_4=0$, the system reduces to the FPUT-$\beta$ model, yielding the following expression for the growth rate
\begin{equation}
\operatorname{Re}[\lambda]=\frac{1}{4}\sqrt{2\zeta^2(3 \beta   \varepsilon -2 \zeta^2)}.
\end{equation}
To determine the transition point, we solve the equation $\operatorname{Re}[\lambda]=0$, yielding the critical value
\begin{equation}
  \zeta_0 = \sqrt{\frac{3\beta \varepsilon}{2}},
\end{equation}
indicating that $\zeta_0$ decreases with decreasing $\varepsilon$. When $\varepsilon$ is sufficiently small, the growth rate of the mode pair corresponding to $m=1$ becomes zero, namely, $\zeta_0 \leq \pi/[2(N+1)]$, with all modes being stable, which leads to the instability threshold $\beta_{c}$ expressed as
\begin{equation}
\beta_{c} = \frac{\pi^2}{6(N+1)^2 \varepsilon}.
\end{equation}
For case I where $\beta=0$, the system simplifies to the Klein-Gordon model. Solving the equation $\operatorname{Re}[\lambda]=0$, the transition curve between the stable and unstable regions in the parameter plane $(\phi_2, \phi_4)$ is given by
\begin{equation}
    \phi_2 = \sec \left(\frac{\pi  q_0}{N+1}\right) \left(1-\frac{3 \varepsilon  }{8\zeta^2}\phi_4\right)+\cos \left(\frac{\pi  q_0}{N+1}\right)-2.
\end{equation}
When $\phi_2 =0$, the instability threshold of $\phi_{4c}$ is
\begin{equation}
    \phi_{4c} = \frac{32 \zeta^2 \sin ^4\left(\frac{\pi  q_0}{2 (N+1)}\right)}{3 \varepsilon}.
\end{equation}
The instability corresponding to the mode pair $m=1$ yields $\zeta = \pi/[2(N+1)]$, the stability threshold becomes
\begin{equation}
    \phi_{4c} =\frac{8\pi^2 }{3 (N+1)^2 \varepsilon} \sin ^4\left[\frac{\pi  q_0}{2 (N+1)}\right].
\end{equation}
The transition curves for case II and case III can be obtained using the same method, which are
\begin{equation}\label{EqPhi2betaA}
\phi_2 = \frac{2 \left(2 \zeta^2-3 \beta  \varepsilon \right)  \sec \left(\frac{\pi  q_0}{N+1}\right)}{\zeta^2}\sin ^4\left[\frac{\pi  q_0}{2 (N+1)}\right],
\end{equation}
and
\begin{equation}\label{EqPhi4betaA}
\phi_4 = \frac{16 \left(2 \zeta^2-3 \beta  \varepsilon \right) }{3 \varepsilon }\sin ^4\left[\frac{\pi  q_0}{2(N+1)}\right],
\end{equation}
respectively.

\section{Results for Modes 5 and 6 in case I}
\label{sec:Mode56}

\begin{figure}[h]
\centering
\includegraphics[width=\linewidth]{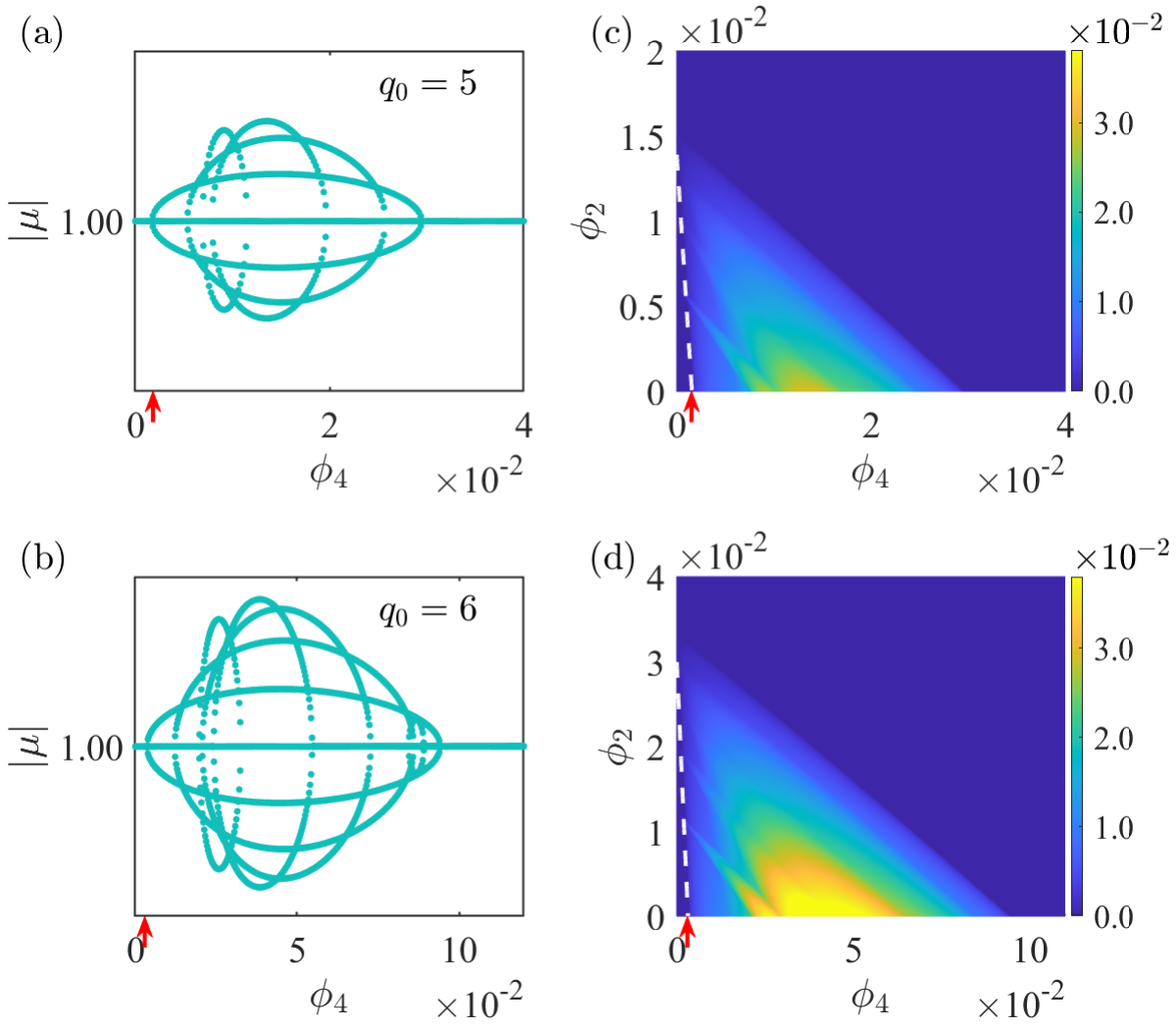}
\caption{Similar to Fig. \ref{fig:localization}, but for the seed modes $q_0=5$ (a, c) and $q_0=6$ (b, d). }
\label{fig:phi4phi2iem5}
\end{figure}

Here we supply more evidence for the verification of the theoretical analysis. Figure \ref{fig:phi4phi2iem5} shows the Floquet multipliers spectrums when $q_0=5$ and $q_0=6$ are chosen as the seed modes in case I. The theoretical prediction given by Eq. (\ref{Eqphi2phi4}) also matches the lower instability boundaries very well, as indicated by the white dashed lines.

\bibliography{Reference.bib,sergejflach.bib}

\end{document}